\documentclass[%
 aip,
 amsmath,amssymb,
 reprint,%
]{revtex4-1}

\usepackage{graphicx}% Include figure files
\usepackage{dcolumn}% Align table columns on decimal point
\usepackage{bm}% bold math
\usepackage[utf8]{inputenc}
\usepackage[T1]{fontenc}
\usepackage{mathptmx}
\usepackage{etoolbox}
\makeatletter
\def\@email#1#2{%
 \endgroup
 \patchcmd{\titleblock@produce}
  {\frontmatter@RRAPformat}
  {\frontmatter@RRAPformat{\produce@RRAP{*#1\href{mailto:#2}{#2}}}\frontmatter@RRAPformat}
  {}{}
}%
\makeatother
\usepackage{sidecap}
\begin{document}
%\emph{•}
%   
\title{Phase- and angle-sensitive terahertz hot-electron bolometric 
plasmonic  detectors based on  FETs with graphene channel and composite 
h-BN/black-P/h-BN gate layer
}
\author{V.~Ryzhii$^{1}$, M. S. Shur$^2$, M.~Ryzhii$^{3}$, V.~Mitin$^4$, C.~Tang$^{1,5}$, and T.~Otsuji$^1$ 
}
\address{
$^1$Research Institute of Electrical Communication,~Tohoku University,~Sendai~ 980-8577, 
Japan\\
$^2$Department of Electrical,~Computer,~and~Systems~Engineering, Rensselaer Polytechnic Institute,~Troy,~New York~12180,~USA\\
$^3$Department of Computer Science and Engineering, University of Aizu, Aizu-Wakamatsu 965-8580, Japan\\
$^4$Department of Electrical Engineering, University at Buffalo, SUNY, Buffalo, New York 14260 USA\\
$^5$Frontier Research Institute for Interdisciplinary Sciences,
Tohoku University, Sendai 980-8578, Japan
}
\begin{abstract}
We propose and analyze  the terahertz (THz)  bolometric vector detectors 
based on the graphene-channel field-effect transistors (GC-FET) with the black-P 
 gate barrier layer or with the composite b-BN/black-P/b-BN gate layer. The phase difference between the signal received by the FET source 
and drain substantially affects the plasmonic resonances. This results in a resonant variation of the detector response on the incoming THz signal phase shift and the THz radiation angle of incidence. 
\end{abstract}

%\begin{IEEEkeywords}
%Graphene, black-Phosphorus, hydrogenated-Boron Nitride,
%terahertz radiation detector,  plasma oscillations, plasma resonance
%\end{IEEEkeywords}
%
\maketitle
%

%\begin{IEEEkeywords}
%Graphene, black-Phosphorus, hexagonal Boron Nitride,
%terahertz radiation detector,  plasma oscillations, plasma resonance
%\end{IEEEkeywords}

The specific properties, in particular,  the band alignment of graphene and the black-P and  black-As (or black-AsP) moderately thick layers~\cite{1,2,3,4} enable the operation of different devices using the thermionic electron or hole emission from the graphene channel (GC) via
the  gate barrier layer (BL). Recently~\cite{5,6}, we predicted 
that the hot-electron bolometric detectors based on the GC-FETs  with
the  GC and the 
black-AsP   BL can exhibit   
fairly high characteristics, for example, responsivity.
The operation of such detectors is associated with the electron heating by the impinging terahertz (THz) radiation stimulating the thermionic electron emission 
 from the GC via the black-AsP  BL into the metal gate (MG). The excitation of the plasmonic oscillations in the  gated GC enables the resonant increase in the electron heating reinforcing  thermionic emission. Plasmonic resonances  can provide  
 elevated detector performance~\cite{5,6}.
 
The strength of the plasmonic resonances depends on the electron collision frequency $\nu$
(determining the plasmonic oscillation damping), i.e., on the electron mobility
in the GC. Since the GC contacting the black-AsP  exhibits a moderate room temperature mobility (for example,~\cite{7}) and not particularly low electron collision frequency,
the GC-FETs with the composite h-BN/black-AsP/h-BN gate BLs
could be more promising.
Indeed,
 the GC-FETs with the GC main part encapsulated in
the h-BN operating at room temperature can provide a fairly pronounced resonant response
due to  high electron mobility and, therefore, a relatively small  $\nu$~\cite{8,9}.
 This challenge can be overcome in
 similar GC-FET bolometric detectors but with the composite incorporating  a relatively narrow black-AsP placed between the 
 h-BN side sections. 
In these detectors, the black-AsP section of the BL serves as the window for the thermionic electron current, while
the rest of the  h-BN can enable a much weaker electron
scattering in the GC. As a result, such GC-FETs 
 can demonstrate a more pronounced resonant response and, hence, elevated responsivity~\cite{10}.

One of the GC-FET detector's feature is a strong influence of the signal electric field
 spatial distribution along the GC on the plasmonic response. This implies that
this distribution  can be varied by changing the phase difference between the signal voltages
created by the antenna system at the GC-FET side contacts (source and drain).

In this paper,   
 we consider the GC-FET bolometric detector structures akin to  those proposed and evaluated previously~\cite{5,6,10}, but focusing on their sensitivity to the THz radiation phase and incident 
 angle. Such a property of THz detectors can be useful for different application (see Ref.~[11] and the references therein).
For the definiteness, we consider
the GC-FET bolometric detectors with the GC and the composite h-BN/black-P/h-BN gate BL.

Figure 1(a)  
 shows the GC-FET 
 structures under consideration. 
The GC-FET comprises the GC sandwiched between the h-BN bottom layer (substrate) and
 the BL consisting of  the lateral h-BN/black-P/h-BN  composite.
 Figure~1(b) shows 
the GC-FET  band diagram at the  BL central section (BL window, $|x| \leq L_C$). Here $2L$ is the length of the GC (approximately equal to the MG length), $2L_C$ is the length of the black-AsP section, and the axis $x$ is directed in the GC plane between the side contacts.
As an alternative, the  WSe$_2$ side sections of the BL   and /or the WSe$_2$ bottom  layer  can be  used
due to the high room temperature mobility~\cite{12} (and, hence, a low electron collision frequency) of the GC containing such a BL.

\begin{figure*}[t]
\centering
\includegraphics[width=13.0cm]{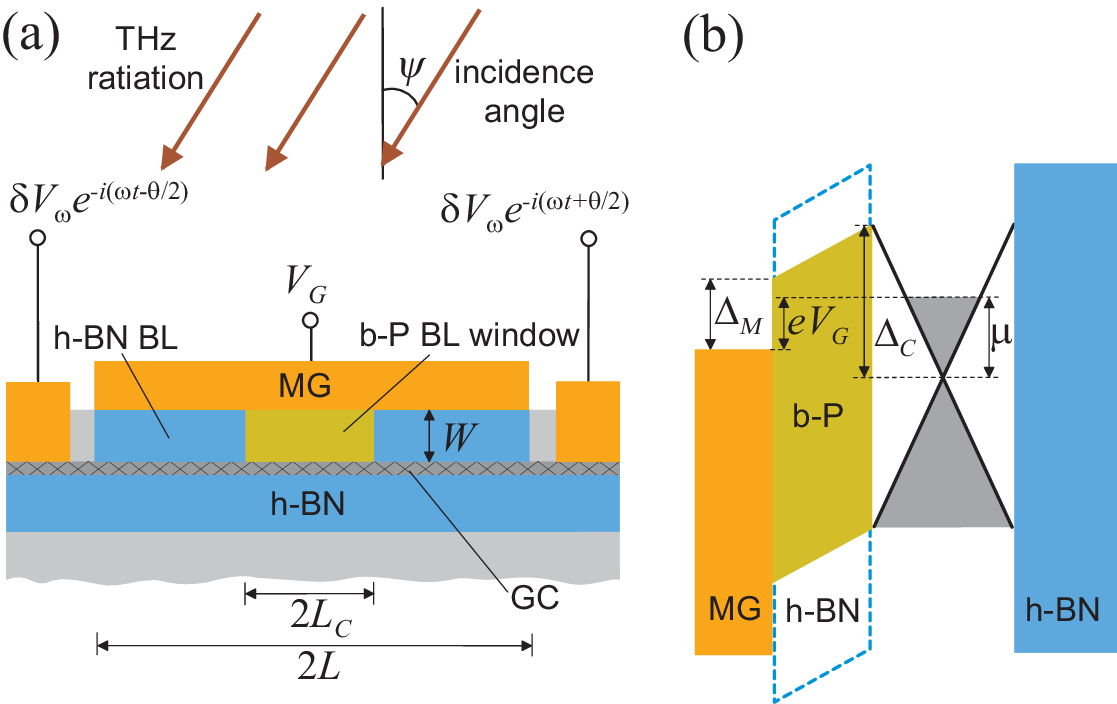}
\caption{Schematic view of the structure of (a) GC-FET bolometric detector
 structure with  
 the composite h-BN/black-P/h-BN BL,  (b) band diagram  
 in the central black-P section of the BL (in the BL window; dashed line corresponds to shape of the h-BN BL in the side regions).}
\label{Fig1}
\end{figure*}

The GC channel is n-doped, so that the electron Fermi energy $\mu \simeq\hbar\,v_W\sqrt{\pi\Sigma_D} \gg T_0$, where $\Sigma_D$ is the donor density in the GC or in the neighboring 
layers, $T_0$ is the lattice temperature, $\hbar$ is the Planck constant and $v_W \simeq 10^8$~cm/s is the electron velocity in graphene.  
The material of the MG is chosen to provide  the condition $\Delta_M = \Delta_C-\mu_D$, where $\Delta_C$ is the difference in the electron affinities  of the Al MG
and black-P BL and the black-P and the GL. For the black-P BL  with  the Al MG,
 $\Delta_M = 85$~meV, $\Delta_C = 225$~meV.
The above condition of the proper band alignment is satisfied at $\mu_D = 140$~meV. This corresponds to $\Sigma_D \simeq 1.6\times 10^{12}$~cm$^{-2}$.
At elevated bias gate voltages $V_G$, the electron Fermi energy in the GC $\mu$ can somewhat exceed $\mu_D$. 

 The main distinction
 of the device under consideration here and that studied  previously is the method
 of the THz radiation input. We assume that the antenna system provides the signal voltages at
 the source and drain contacts shifted by the phase $\theta$. 
 According to the latter, the signal component  GC potential $\delta \varphi_{\omega}$
 with respect to the MG potential
 obeys the following conditions at the edges of the source ($x=-L$) and  drain ($x=L$) contacts:
$\delta \varphi_{\omega}|_{\pm L} = \delta V_{\omega}\exp(-i\omega t \pm i\theta/2)$. 
 Here $\delta V_{\omega}$ and $\omega$ are the amplitude and the frequency  of the electric signal provided by the antenna system under the effect of the impinging THz radiation, the axis $x$ is directed in the GC plane from the source to the drain, and $2L$
 is the spacing between these contacts(approximately equal to the GC length).

 The signal voltages at the  source and the drain excite the ac electric field  $\delta E_{\omega}$ in the GC causing the absorption of the signal energy by the electrons.
 The absorbed power per unit of the GC area averaged over the radiation period (the averaging is denoted by the symbol $\langle...\rangle$)
 is equal to the Joule power $Q_{\omega} = {\rm Re} \sigma_{\omega}\langle\delta E_{\omega}|^2\rangle$,
 where $\sigma_{\omega}$ is the GC ac Drude conductivity.
 The radiation absorption results in the variation, $\langle\delta T_{\omega}\rangle$, of the effective  electron temperature in the GC. The latter leads to the variation of the thermionic
 currents from the GC via the BL into the MG ~\cite{5,6}:

\begin{eqnarray}\label{eq1}
\langle \delta J_{\omega}\rangle^C =  Hj^{max} F
\int_{-L_C}^{+L_C}dx\frac{\langle \delta T_{\omega}\rangle}{T_0}.
\end{eqnarray} 
Here 
$F= (\Delta_M/T_0)\exp(-\Delta_M/T_0)$ is "the thermionic emission factor",
$j^{max} = e\Sigma_D/\tau_{\bot}$ is the maximal value of the current density from the GC to the MG, $\tau_{\bot}$ is the  electron
try-to-escape  time, and $H$ is the GC  width.

In the following, we consider  the GC-FETs with $L_C \ll L$. In these GC-FETs, the plasmonic oscillations resonances   in the whole gated GC are primarily attenuated by  the electron scattering frequency $\nu$ in the GC encapsulated in the h-BN, which can be fairly small~\cite{9,10}, while the effect of a narrow black-AsP window section is weak.

For the axis $x$ spatial distribution of  the Joule power associated with the
ac electric field produced by the plasmonic oscillations 
 in the gated GC~\cite{13,14,15}, one can obtain

\begin{eqnarray}\label{eq2}
%{\rm Re}\sigma_{\omega}\langle |\delta E_{\omega}|^2\rangle
\langle Q_{\omega}\rangle = \frac{\sigma_0}{2}\biggl(\frac{\delta V_{\omega}}{L}\biggr)^2\biggl(\frac{\pi\nu}{2\Omega_P}\biggr)^2\frac{\omega}{\sqrt{\nu^2+\omega^2}}\nonumber\\
\times\biggl|
\biggl[\frac{\cos(\gamma_{\omega}x/L)\sin(\theta/2)}{\sin \gamma_{\omega}} -
\frac{\sin(\gamma_{\omega}x/L)\cos(\theta/2)}{\cos \gamma_{\omega}}\biggr]\biggr|^2.
\end{eqnarray}
Here $\gamma_{\omega} = \pi\sqrt{\omega(\omega+i\nu)}/2\Omega_P$ and $\Omega_P = (\pi\,e/\hbar\,L)\sqrt{\mu_DW/\kappa}$ is the plasmonic frequency 
with $W$ and $\kappa$ being the thickness and dielectric constant of the h-BN/b-P/h-BN BL, respectively.
The right-hand side of Eq.~(2)describes the resonant overshoots at $\omega \simeq \Omega_P$ and at higher plasmonic resonances.

\begin{figure}[t]
\centering
\includegraphics[width=7.0cm]{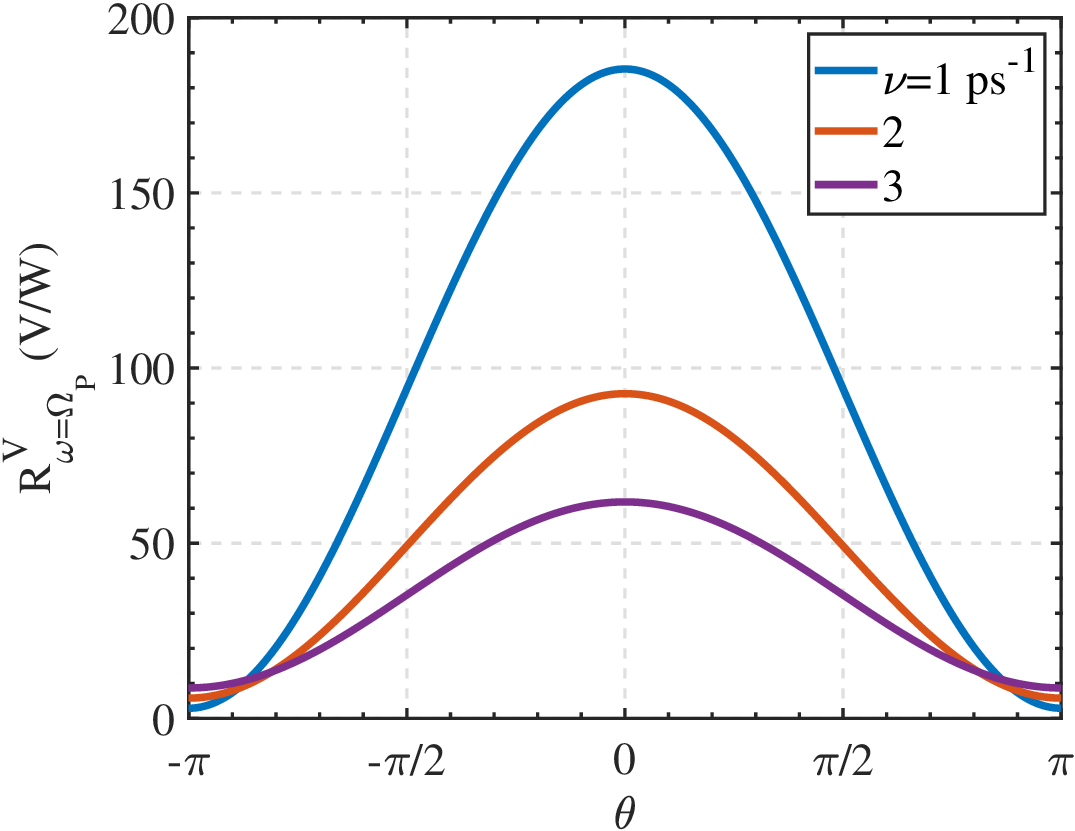}
\caption{The GC-FET detector resonant responsivity $R_{\omega = \Omega_P}^V$ as  function of phase difference $\theta$ at different values of collision frequencies $\nu$.
}
\label{Fig2}
\end{figure}

The spatial distribution 
of
$ \langle\delta T_{\omega}\rangle$ is described by  the one-dimensional 
electron heat transport equation, which for the device structure under consideration is presented in the following form~\cite{5,6}:

\begin{eqnarray}\label{eq3}
-h\frac{d^2 \langle\delta T_{\omega}\rangle}{d x^2}+ 
\frac{ \langle\delta T_{\omega}\rangle}{\tau_{\varepsilon}}
=\langle Q_{\omega}\rangle.
\end{eqnarray} 
Here 
$h \simeq v_W^2/2\nu$ is the  electron thermal conductivity in the GC per electron
(this  corresponds to the Wiedemann-Franz relation),  $v_W \simeq 10^8$~cm/s is the characteristic electron velocity in GCs, $\tau_{\varepsilon}$ is the electron energy relaxation time in the GC.
Equation~(3) describes the electron heat  propagation along the GC and accounts for the electron heat transfer to the source and drain contacts.
The latter can be significant due to high values of the electron heat conductivity~\cite{16,17}.
The second term in the left-hand side  of Eq.~(3)
reflects the energy loss by the GC due to its transfer to the lattice~\cite{18,19,20,21,22}. The term describing the electron thermal energy leakage via the BL~\cite{5,6,23} is disregarded because of the small length, $2L_C$, of the window
section~\cite{11}. Equation~(3) is supplemented by the boundary conditions:
 $<\delta T_{\omega}>|_{x = \pm L} =0$. 

Since the ac electric field along the GC is not a particularly strongly varying function of the coordinate $x$, one can replace the right-hand side of Eq.~(3) by its spatially averaged value $\overline {\langle Q_{\omega}\rangle}$ (the Joule power
averaged over the GC).
Considering the signal frequencies close to the frequency of the fundamental
 plasmonic resonance and that the quality factor of this resonance$(2\Omega_P/\pi\nu)^2 \gg 1$, we obtain  

\begin{eqnarray}\label{eq4}
\overline {Q_{\omega}}  \simeq \sigma_0\biggl(\frac{\delta V_{\omega}}{L}\biggr)^2
%\nonumber\\
\biggl[\cos^2\biggl(\frac{\theta}{2}\biggr)
+ \biggl(\frac{\pi\nu}{4\Omega_P}\biggl)^2\sin^2\biggl(\frac{\theta}{2}\biggr)\biggr]
\end{eqnarray}
Using Eq.~(4), we arrive at the approximate solution of Eq.~(3) with the above boundary conditions in the form:
 
\begin{eqnarray}\label{eq5}
<\delta T_{\omega}>\simeq \overline {Q_{\omega}} \tau_{\varepsilon}
\biggl[1 - \frac{\cosh(x/{\mathcal L})}{\cosh(L/{\mathcal L})}\biggr],
\end{eqnarray} 
where
${\mathcal L} = \sqrt{h\tau_{\varepsilon}}
= v_W\sqrt{\tau_{\varepsilon}/2\nu}$
is the electron cooling length.
Equation~(5) yields

\begin{figure}[b]
\centering
\includegraphics[width=8.0cm]{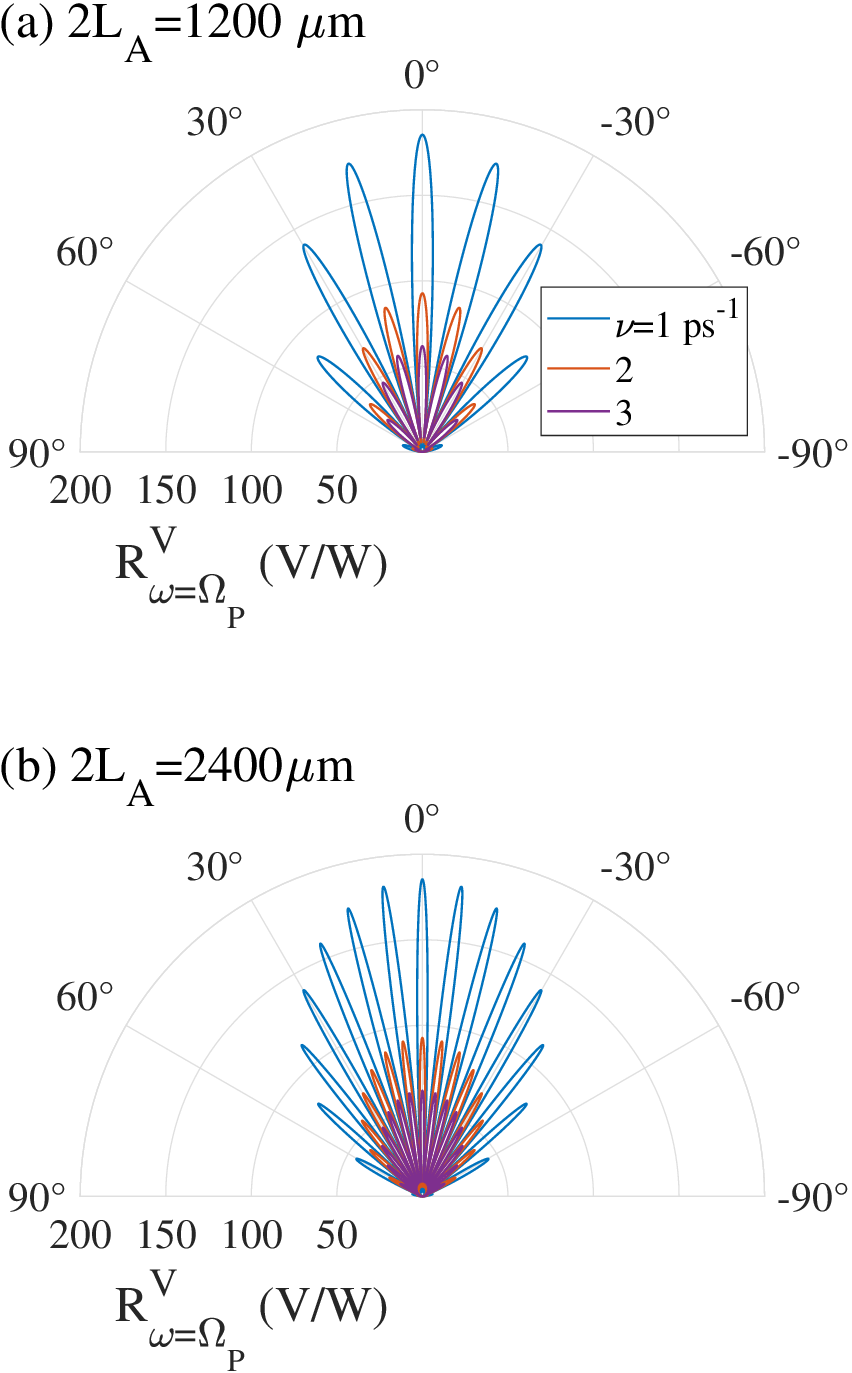}
\vspace{+5mm}
\caption{The GC-FET detector resonant responsivity $R_{\omega = \Omega_P}^V$ as  a function of  the angle of incidence $\psi$ for different collision frequencies $\nu$ for different spacing between the antennas (a)  $2L_A = ~1200~\mu$m and (b)
  $2L_A = ~2400~\mu$m. }
\label{Fig3}
\end{figure}

\begin{eqnarray}\label{eq6}
\int_{-L_C}^{L_C}dx \langle\delta T_{\omega}\rangle \simeq 2\overline {Q_{\omega}} \tau_{\varepsilon}
\biggl[L_C - {\mathcal L}\frac{\sinh(L_C/{\mathcal L})}{\cosh(L/{\mathcal L})}\biggr]
\nonumber\\
\simeq
2L_C\overline {Q_{\omega}} \tau_{\varepsilon}
\biggl[1 - \frac{1}{\cosh(L/{\mathcal L})}\biggr].
\end{eqnarray} 

\begin{table}[]
\centering
\caption{\label{table} GC-FET parameters} \vspace{2 mm}
\begin{tabular}{|r|c|}
\hline
GC length, $2L$&   $1.5~\mu$m \\
\hline
Gate layer thickness, $W$& 17.4~nm\\
\hline
Electron Fermi energy, $\mu$& 140~meV\\
\hline
Plasmonic frequency, $\Omega_P/2\pi$& 1.0 THz\\
\hline

Lattice temperature, $T_0$& 25 meV\\
\hline
Energy relaxation time, $\tau_{\varepsilon}$& 10~ps\\
%\hline
%Try-to-escape time, $\tau_{\bot}$& 10~ps\\
\hline
Antenna gain, $g$& 1.64\\
\hline
\end{tabular}
%\caption{\label{table} Parameters of the GC-FET structures under consideration.} 
\end{table}

The GC-FET bolometric detector voltage responsivity
 at the fundamental plasmonic resonance is given by
 $R_{\omega = \Omega_P}^V = \langle \delta J_{\Omega_P}/S_{\Omega_P})r_L  $, where $S_{\Omega_P}$ is 
 the power of the normally impinging THz radiation  received by the antennas and $r_{Load}$ is the load resistance assumed to be equal to the GC-MG resistance.
The quantity  $(\delta V_{\omega})^2 \propto S_{\omega} \cos^2\psi $
   (see, for example,~\cite{24}), where $\psi$ is the angle of the radiation incidence. In the case of the normal incidence ($\psi =0$), using Eqs.~(1) and (5) with Eq.~(3),  we 
  arrive at the following formula: 

\begin{eqnarray}\label{eq7}
R_{\omega = \Omega_P}^V \simeq  R^V\biggl[\cos^2\biggl(\frac{\theta}{2}\biggr)
+ \biggl(\frac{\pi\nu}{4\Omega_P}\biggl)^2\sin^2\biggl(\frac{\theta}{2}\biggr)\biggr].
\end{eqnarray}
Here \begin{eqnarray}\label{eq8}
R^V = 
\frac{32}{137g}\biggl(\frac{\hbar}{eT_0}\biggr)
\biggl(\frac{\Delta_M}{\mu}\biggr)
\biggl(\frac{v_W^2\tau_{\varepsilon}}{L^2\nu}\biggr)\biggl[1 - \frac{1}{\cosh(L/{\mathcal L})}\biggr].
\end{eqnarray}
\begin{figure}[t]
\centering
\includegraphics[width=8.0cm]{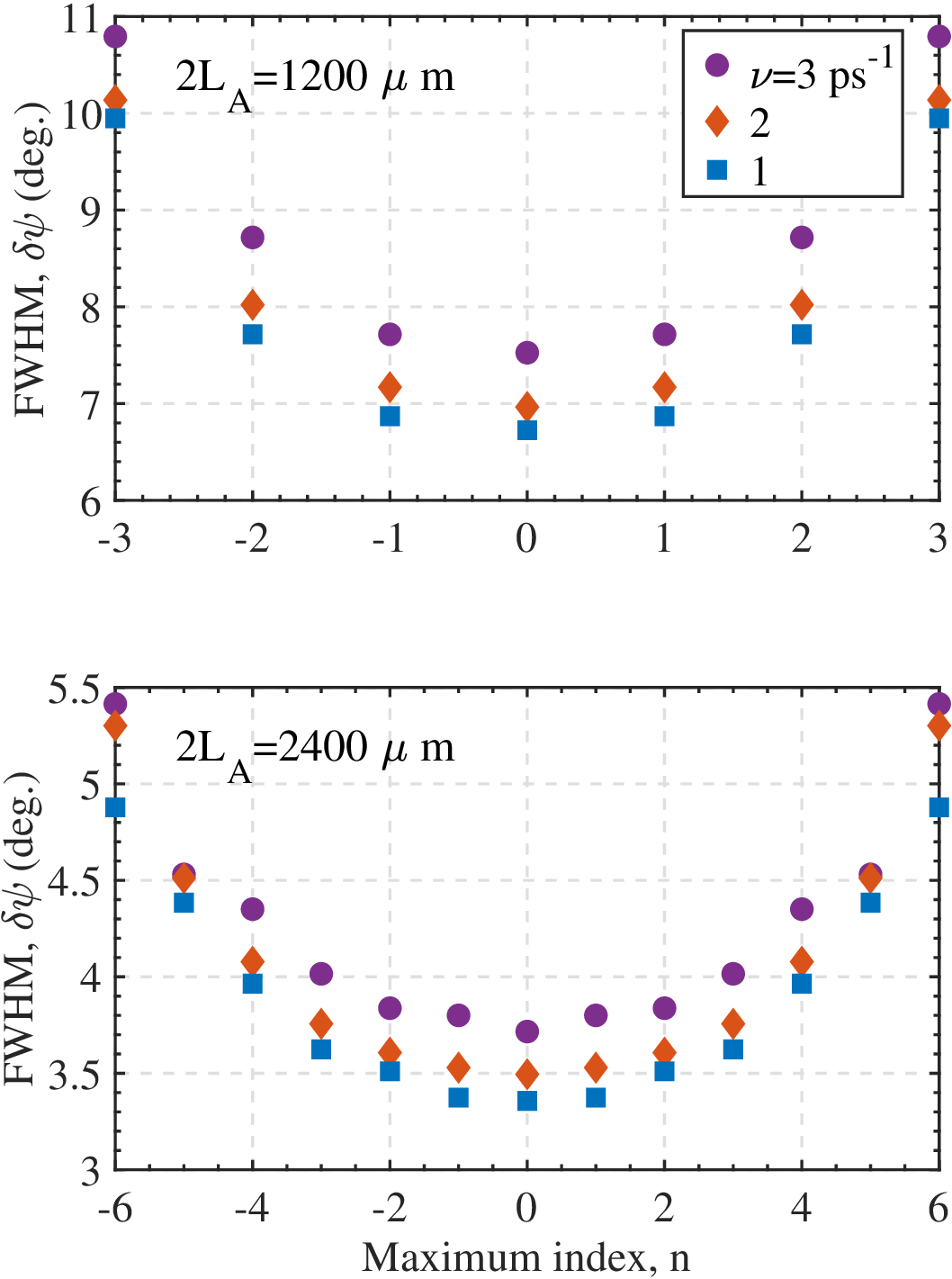}
\caption{
Angular FWHM of the different peaks of resonant responsivity $R_{\omega = \Omega_P}^V$ corresponding to Fig.3: upper panel -  $2L_A = ~1200~\mu$m and lower panel - 
  $2L_A = ~2400~\mu$m.  }
\label{Fig4}
\end{figure}
It is worth noting that $R^V$ is independent of $\tau_{\bot}$. This is because
the GC- to MG THz radiation stimulated current  is proportional to $\tau_{\bot}^{-1 }$, while the GC-MG resistance (and, hence, the chosen load resistance) is proportional to $\tau_{\bot}$.

Figure~2 shows the phase dependence of the resonant voltage responsivity $R_{\omega = \Omega_P}^V$ calculated for the GC-FET bolometric detectors with the band parameters  indicated above and 
characterized by different values of the electron collision frequency $\nu$, i.e.,
different electron mobilities $M$.
Additional parameters are listed in Table I.
One can see that  the resonant voltage responsivity can be fairly high at
certain values of the phase difference $\theta$, especially in the GC-FETs with
relatively small collision frequencies $\nu$ (compare, for example,  with Refs.~[25,26]). The electron mobility in the GC with the values of $\nu$ used in Fig.~2 can be estimated as $M \simeq (2.4 - 7.1)\times 10^4$~cm$^2$/Vs, which is reasonable for GCs encapsulated in n-BN layers.

If the phase shift is associated with the inclined radiation incidence,
its value is equal to $\theta = \omega\,2L_A\sin\psi/c = 4\pi\,L_A\sin\psi/\lambda_{\omega}$, where $2L_A$ is the distance between the antennas producing the voltage signals $\delta V_{\omega}\exp[-i(\omega t \mp\theta/2)]$ [not shown in Fig.~1(a)]
 at the source and drain, respectively, $c$ is the speed of light in vacuum, and
 $\lambda_{\omega}$ is the THz radiation wavelength. Then we obtain 
\begin{eqnarray}\label{eq9}
R_{\omega = \Omega_P}^V \simeq  
 R^V\cos^2\psi\nonumber\\
 \times\biggr[\cos^2(l_A\sin\psi)
+ \biggl(\frac{\pi\nu}{4\Omega_P}\biggl)^2\sin^2(l_A\sin\psi)\biggr],
\end{eqnarray}
where $l_A = L_A\Omega_P/c$ is proportional to the ratio of the antenna spacing and the impinging THz radiation resonant 
wavelength and $g$ is the antenna gain. The phase shift can be also associated
with the circular polarization of the incident THz radiation~\cite{27}.

Figure~3 demonstrates the angular dependence of the resonant responsivity  $R_{\omega} = \Omega_P^V$ 
of the detectors with different spacing, $L_A$, between the antennas
calculated for different values of collision frequency $\nu$.
As seen in Fig.~3, the angular width of the responsivity peaks is fairly small.
This is confirmed by the values of the peaks full width of half maximum (FWHM) shown in Fig.4. 

Due to the voltage dependence of the plasmonic frequency $\Omega_P$ (associated with the variation of the Fermi energy), the plasmonic resonance can be voltage tuned. This might lead to the controllability of the responsivity phase and incidence angle.

In conclusion,  we have proposed the THz GC-FET bolometric detectors with the  h-BN/black-P/h-BN BLs and predicted that these detectors
can exhibit fairly high room temperature responsivity
 with
a pronounced phase and angular selectivity. Since the detector response strongly depends on the
plasmonic resonance (i.e., the ratio $\omega/\Omega_P$), the latter is also very sensitive to the impinging radiation
frequency.

The work at TU, UoA, and UB was supported by the Japan Society for Promotion of Science (KAKENHI  Nos. 21H04546, 20K20349),
Japan and the RIEC Nation-Wide Collaborative Research
Project No. R04/A10, Japan.  The work at RPI was supported by AFOSR (contract number FA9550-19-1-0355).

\end{document}